\documentclass[fleqn,usenatbib]{mnras}

\usepackage{newtxtext,newtxmath}

\usepackage[T1]{fontenc}

\DeclareRobustCommand{\VAN}[3]{#2}
\let\VANthebibliography\thebibliography
\def\thebibliography{\DeclareRobustCommand{\VAN}[3]{##3}\VANthebibliography}

\usepackage{graphicx}	
\usepackage{amsmath}	

\usepackage{soul} 

\title[Sloshing Spiral in A2107]{Origin of the Sloshing Spiral in the Abell 2107 Galaxy Cluster}

\author[M. K. Erdim]{
M. Kıyami Erdim,$^{1}$\thanks{E-mail: mkiyami@yildiz.edu.tr (MKE)}
\\
$^{1}$Physics Department, Yıldız Technical University, İstanbul, Türkiye\\
}

\date{Accepted XXX. Received YYY; in original form ZZZ}

\pubyear{\the\year{}}

\begin{document}
\label{firstpage}
\pagerange{\pageref{firstpage}--\pageref{lastpage}}
\maketitle

\begin{abstract}

We present a detailed thermodynamic analysis of a sloshing spiral discovered in the intracluster medium (ICM) of the galaxy cluster Abell 2107, based on deep Chandra X-ray Observatory observations. Spectral analysis reveals that the spiral arm is significantly cooler and exhibits lower entropy and higher metallicity compared to the surrounding ICM, consistent with an origin in the cool, dense cluster core. We detect three cold fronts at the edges of the spiral, roughly aligned along an east-west axis. This alignment is consistent with predictions from numerical simulations of off-axis merger events. Subregion analysis of the spiral shows that the coldest and lowest-entropy gas is concentrated in the central part of the arm, while the side regions show evidence of mixing and interaction with the ambient medium. Among the examined thermodynamic quantities, entropy provides the most robust contrast between the sloshing gas and its surroundings. Notably, no optical or X-ray counterpart to a potential perturber is detected, suggesting that the responsible perturber may currently be below the detection limit or otherwise difficult to identify, while a dark matter-dominated or dark matter-only subhalo remains one possible scenario. Our findings provide new insights into the dynamical history of A2107 and demonstrate the importance of deep X-ray exposures for revealing subtle ICM structures in apparently relaxed clusters.
\end{abstract}




\section{Introduction} \label{sec:introduction}

Galaxy clusters, the most massive virialized structures in the universe, evolve through a process of hierarchical merging and accretion. These merger events, ranging from major collisions to minor group accretions, leave distinct signatures in the physical properties of the intracluster medium (ICM). The discovery of many of these features became possible with the high angular resolution X-ray observations of the \textit{Chandra} Observatory.

In many clusters, radiative cooling of the ICM over cosmic time leads to the formation of dense, low-entropy cool cores characterized by centrally peaked X-ray emission and short cooling times \citep{fabian1994cooling}. Such cool cores are dynamically sensitive to gravitational perturbations, and merger events can displace the central low-entropy gas from the cluster center, causing it to oscillate within the gravitational potential well. These oscillations produce sharp surface brightness and density discontinuities called cold fronts, which are contact discontinuities separating gas phases where the denser side is cooler while the thermal pressure remains approximately continuous \citep{markevitch2000chandra,vikhlinin2001moving,ascasibar2006origin,markevitch2007shocks}. The gas on the denser side typically exhibits lower entropy and often enhanced metallicity compared to the surrounding ICM \citep[see for extensive review][]{markevitch2007shocks}. 

When the merger occurs with a non-zero impact parameter (i.e., off-axis), the infalling subhalo transfers angular momentum to this central cool core gas, displacing it from the potential minimum and initiating a sloshing motion. Simulations show that this mechanism leads to the development of characteristic spiral-like cold front patterns \citep{ascasibar2006origin,zuhone2010stirring}. Consistent with these predictions, spiral structures have been identified in a number of X-ray studies of galaxy clusters, including Perseus, Centaurus, Virgo, A2029, A496, A2204, A795 and others \citep[e.g.,][]{churazov2003xmm,lagana2010spiral,clarke2004complex,ghizzardi2014metal,kadam2024sloshing_2,zheng2025imaging}.

Depending on the mass ratio of the merging halos and the impact parameter of the encounter, a wide range of ICM distortions can arise. While some mergers produce pronounced morphological disturbances and easily detectable X-ray counterparts, others leave only subtle signatures with no obvious optical or X-ray perturber identifiable \citep{dupke2007different,paterno2013deep}. Such observations could reflect the limitations of current instruments. However, numerical simulations offer an alternative explanation for these observations: even dark matter-only halos or gas-poor subhalos, with little or no baryonic matter, can induce sloshing features in the cluster core through purely gravitational perturbations \citep{ascasibar2006origin,zuhone2010stirring}. Thus, sloshing spirals and cold fronts could arise from perturbers that remain invisible to the observations.

Despite their widespread detection, our understanding of sloshing spirals' role in regulating cluster thermal evolution remains incomplete. Early studies suggested that sloshing could efficiently mix hot gas from the outskirts with the dense, cool core, thereby quenching catastrophic cooling flows \citep{markevitch2007shocks}. However, recent observations and simulations indicate that this mixing is far less efficient than initially thought \citep{ghizzardi2014metal, zuhone2011sloshing}. The suppression of mixing has been attributed to magnetic fields that are "draped" along the cold front surfaces, which can inhibit Kelvin-Helmholtz instabilities and suppress thermal conduction \citep{chen2017gas, zuhone2011sloshing}. Nevertheless, edge distortions consistent with Kelvin-Helmholtz instabilities have been reported at spiral boundaries, offering indirect probes of ICM viscosity and magnetic field geometry \citep{zuhone2013cold, roediger2013kelvin}. Consequently, while sloshing may not be the dominant mechanism for regulating cooling, it likely contributes to the overall dynamics of the cluster core on larger scales, potentially complementing AGN feedback in maintaining the thermal balance of the ICM \citep{ghizzardi2014metal, li2025x}. These considerations underscore the need for detailed observational investigations of sloshing features.

Understanding the role of dynamical perturbations in cluster cores requires detailed measurements of thermodynamic properties across morphological disturbances such as sloshing spirals, cold fronts, and shocks. By examining temperature, entropy, and metallicity variations coupled with high-resolution imaging of small-scale morphology, we can characterize how gas of different phases interacts and mixes within the ICM. Such detailed studies provide crucial insights into the underlying physical processes, including magnetic field configurations, plasma instabilities, and heat transport mechanisms. Clusters hosting prominent dynamical structures therefore serve as natural laboratories for probing the complex interplay between gas dynamics, magnetic fields, and feedback processes in shaping the thermal evolution of galaxy cluster cores.

Abell 2107 provides a valuable system for investigating these processes. Previous studies have characterized A2107 as a remarkably regular and relaxed cluster \citep{girardi1997optical,kalinkov2005rotation}, with symmetric X-ray morphology suggesting a quiescent state \citep{lagana2010spiral,song2018redshift,li2025x}. However, through deep Chandra observations, we identify a prominent sloshing spiral and multiple cold fronts along the spiral boundaries. Notably, no obvious perturber is detectable in optical or X-ray observations, suggesting that the perturber may lie below the current detection limit, while a dark matter-dominated or dark matter-only subhalo remains a possible alternative scenario. A2107 therefore offers an excellent opportunity for a detailed investigation of the thermodynamic structure and internal properties of cluster-scale dynamical disturbances.

In this work, we report the discovery of a sloshing spiral in the galaxy cluster A2107 using deep \textit{Chandra} observations. We analyze its thermodynamic properties and characterize the multiple cold fronts identified along the edges of this spiral. Finally, we discuss the possible formation scenarios responsible for these features. 

Throughout this paper, we adopt a flat $\Lambda$CDM cosmology with $H_0 = 70~\mathrm{km~s^{-1}~Mpc^{-1}}$, $\Omega_\mathrm{m} = 0.3$, and $\Omega_\Lambda = 0.7$, and all quoted uncertainties correspond to the $1\sigma$ confidence level, unless otherwise stated. The paper is structured as follows: Section \ref{sec:analysis} describes the data reduction and analysis, Section \ref{sec:results} presents the results, Section \ref{sec:discussion} discusses the implications, and Section \ref{sec:conclusion} summarizes the conclusions.

\section{Data Reduction and Analysis} \label{sec:analysis}

The Abell 2107 galaxy cluster (RA: 15$^{h}$39$^{m}$38.4$^{s}$, Dec: +21$^{d}$47$^{m}$20.0$^{s}$) was observed by Chandra in seven separate observations conducted between 2004 and 2022, yielding a total raw exposure time of approximately $\sim$144 ks (see Table\ref{tab:obs}). All the data were processed using \textit{CIAO} (ver. 4.16) with the latest calibration files from \textit{CALDB} (ver. 4.11).

\begin{table}
   \centering
   \renewcommand{\arraystretch}{1.2} 
   \begin{tabular}{lccc}
      \hline \hline
      Obs ID & t$_{raw}$ (ks) & t$_{filt}$ (ks) & Date \\\hline
      4960   & 35.57          & 35.15           & 2004 \\
      24350  & 14.88          & 14.82           & 2021 \\
      24351  & 16.85          & 16.66           & 2021 \\
      26152  & 14.88          & 14.88           & 2021 \\
      26153  & 12.47          & 12.47           & 2021 \\
      23851  & 24.76          & 24.76           & 2022 \\
      26443  & 24.76          & 24.76           & 2022 \\ \hline
      Total: &                & $\sim$143       &      \\ \hline
   \end{tabular}
   \caption{Information on the \textit{Chandra} observations of Abell 2107.}
   \label{tab:obs}
\end{table}

\subsection{Data Preparation} 

We present a summary of the data reduction procedure. Initially, each observation was reprocessed using the \textit{chandra$\_$repro} script to apply the latest calibration files. Flare contamination from high-energy particle events was removed using the \textit{deflare} tool, generating Good Time Interval (\textit{GTI}) files, and only these uncontaminated periods were retained in the final event files. Point sources were identified and removed with the \textit{wavdetect} tool to ensure that the analysis focused solely on the cluster's diffuse emission. After excluding point sources, the gaps left by the removed sources were filled using the \textit{dmfilth} tool; however, this step was applied only to image files for imaging purposes and did not affect spectral analyses. For background modeling, a local background approach was adopted, selecting low-emission regions far from the cluster center in the field of view (an annulus with radii of $\sim$298.5-383.8 kpc, concentric with the cluster center). Finally, the processed event and image files were used for spectral extraction and imaging analyses.

\begin{figure*}
   \centering
   \includegraphics[width=\textwidth, trim=0 0 0 0, clip]{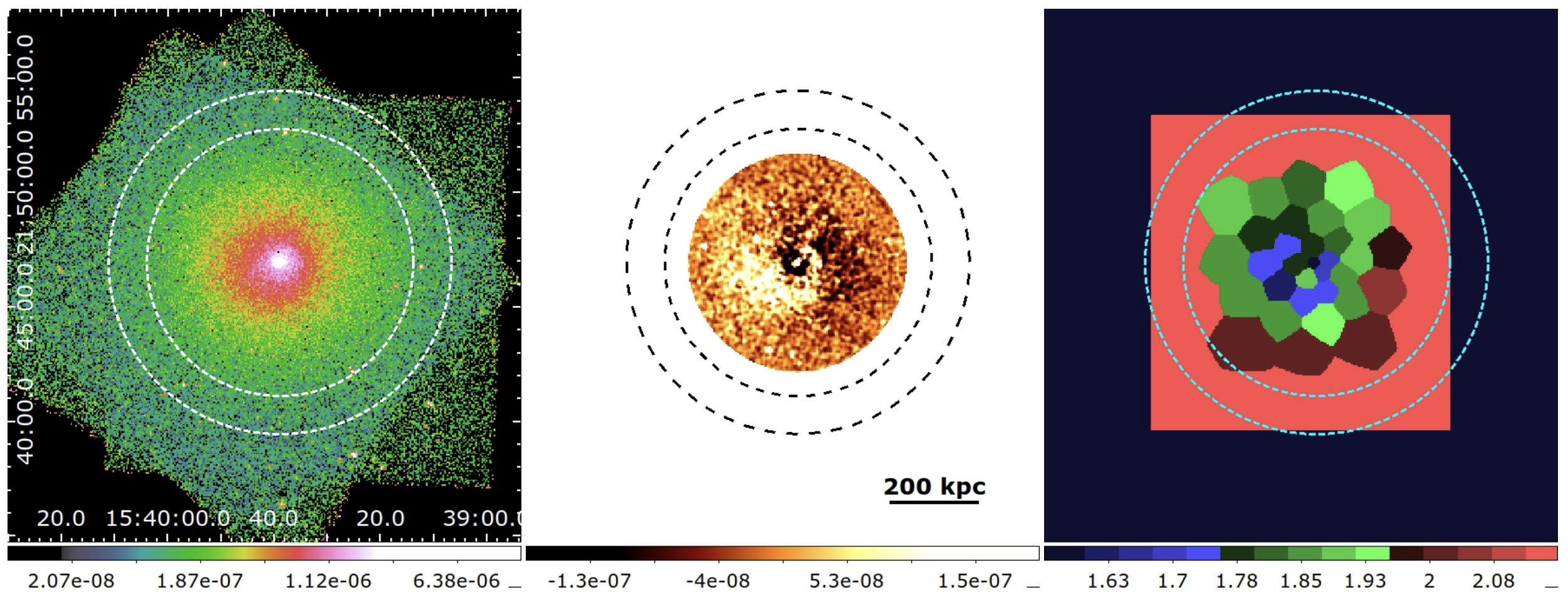}
   \caption{Merged X-ray image (\textit{left}), double elliptical $\beta$-model subtracted residual image (\textit{middle}), and Pseudo-temperature map (\textit{right}) of A2107. The dashed annulus indicates the background extraction region.}
   \label{fig:A2107-xray-cavity-vorbin}
\end{figure*}

\subsection{Spectral Analysis}

Spectral analysis was subsequently performed to quantitatively examine the thermodynamic properties across these morphological regions of the ICM. Spectral extraction for each region of interest was carried out using the \textit{specextract} tool separately for each observation, producing source and background spectra together with the corresponding response files (RMFs and ARFs). $\chi^2$ statistics were used, with the spectra binned to a minimum of 15 photons per bin. All spectra were then fitted simultaneously using \textit{Sherpa} \citep[v4.16]{freeman2001sherpa} with each dataset loaded as an independent data group. An absorbed (\textit{PHABS}; \citealt{balucinska1992photoelectric}) thermal emission model (\textit{APEC}; \citealt{smith2001collisional}) was adopted to represent the plasma emission, within the 0.7-7.0 keV energy range. The temperature, abundance, and normalization parameters were linked across all data groups, while the Galactic column density ($N_{H} = 0.0458$; \citealt{bekhti2016hi4pi}) and the redshift ($z = 0.0414$) were kept fixed. Elemental abundances were calculated relative to Solar values using the LPGS table \citep{lodders2009abundances}. Based on the results of these simultaneous fits, temperature, metallicity, pseudo-density, pseudo-pressure and pseudo-entropy values were calculated for each region, enabling a detailed thermodynamic comparison of the spiral-like substructure with its surrounding environment.

Within the scope of this study, relative spatial variations of thermodynamic parameters are sufficient to characterize substructures. Therefore pseudo-values are used for density, pressure, and entropy, rather than their actual values. The emission measure values ($\mathrm{EM}$; Eq. \ref{eq:emission_measure}) were obtained by dividing the normalizations ($\mathcal{N}$; Eq. \ref{eq:apec_norm}) by the area of the spectral regions, which is proportional to the square of the electron density (${n_e}^2$). Therefore square root of $\mathrm{EM}$ used as pseudo-density. Subsequently, pseudo-pressure and pseudo-entropy were calculated using the relations in Eq. \ref{eq:pressure_entropy}.

\begin{equation}
   \mathrm{EM} =
   \mathcal{N} / \mathrm{Area}
   \label{eq:emission_measure}
\end{equation}

\begin{equation}
   \mathcal{N} =
   \frac{10^{-14}}{4\pi \left[D_A (1+z)\right]^2}
   \int n_{\mathrm{e}}\, n_{\mathrm{H}} \, dV
   \label{eq:apec_norm}
\end{equation}

\begin{equation}
   \begin{split}
      P &= n k T        &\quad &\Rightarrow \quad &P \propto &k T (\mathrm{EM})^{1/2}  \\
      S &= k T n^{-2/3} &\quad &\Rightarrow \quad &S \propto &k T (\mathrm{EM})^{-1/3}
   \end{split}
   \label{eq:pressure_entropy}
\end{equation}

\subsection{Imaging Analysis}

High spatial resolution of the \textit{Chandra} observatory provides opportunity to inspect subtle visual features of the cluster. Initially we created pseudo-temperature map, adopting the hardness-ratio (HR) approximation method, in order to search for temperature variations in the ICM. Followingly, $\beta$-model subtracted residual image was created for revealing possible X-ray cavities or areas with excess emission (i.e. deviations from spherical symmetry). Finally, radial surface brightness profiles were extracted for identifying and modeling density discontinuies. Procedure of the adopted methods are summarized below.

\subsubsection{Hardness-Ratio Map}
Hardness-ratio (HR) maps are useful tools when calculating relative temperature variations in the ICM is adequate, instead of the actual values. Within the scope of this study, these relative temperature variations were used to inspect the general temperature structure of the ICM, before proceeding further analysis and spectral confirmation. Firstly, background subtracted and exposure corrected merged image of the cluster was binned with Weighted Voronoi Tessellation method using \texttt{vorbin} tool \citep{cappellari2003adaptive}, which produces voronoi shaped bins with similar signal-to-noise ratios. Then the soft-band (0.7-1.6 keV) and the hard-band (1.6-7.0 keV) images were binned using these defined voronoi regions in the previous step. Finally hard-band and soft-band images were divided to obtain HR map. Therefore, each voronoi bin shows the hardness ratio (i.e. pseudo-temperature) of the contained pixels. The HR map is presented in the right panel of Figure \ref{fig:A2107-xray-cavity-vorbin}.

\subsubsection{Residual Image}
Residual image is created by modeling the ICM emission with 2D double elliptical $\beta$-model and subtracting the best-fit model from the data. Thus, the obtained residual image shows the deviations from the symmetric model. The model consists of two elliptical $\beta$-model components: one describing the main cluster emission extending to large radii, and the other representing the centrally peaked emission of the cool core. Additionally a constant background model was added for non-ICM emissions.

\subsubsection{Radial Surface Brightness Profiles}
Detecting density discontinuities in the ICM emission are crucial for confirming the existance of dynamical features (e.g. shocks, cold fronts). Radial surface brightness profiles are indicators of these discontinuities. Profiles were extracted and modelled by \texttt{pyproffit} tool \citep{eckert2020low}. Density jumps of the discontinuities are modeled by broken-powerlaw.

This package models the discontinuity by utilizing Eq. \ref{eq:proffit_1} and Eq. \ref{eq:proffit_2}. Here, $r_f$ denotes the radius of the discontinuity. The parameters $\alpha_1$ and $\alpha_2$ represent the powerlaw indices inside and outside $r_f$, respectively, while $C$ is the jump ratio across the discontinuity. (See \cite{eckert2020low} for further details on the methods and models used by the \texttt{pyproffit} package.) 

\begin{equation}
   I(r) = I_0 \int F(\omega)^2 \, d\ell + B \quad , \quad (\omega^2 = r^2 + \ell^2)
   \label{eq:proffit_1}
\end{equation}

\begin{equation}
   F(\omega) =
   \begin{cases}
      \omega^{-\alpha_1},               & \omega < r_f,  \\
      \dfrac{1}{C}\,\omega^{-\alpha_2}, & \omega \ge r_f
   \end{cases}
   \label{eq:proffit_2}
\end{equation}

\begin{figure*}
   \centering
   \includegraphics[width=\textwidth]{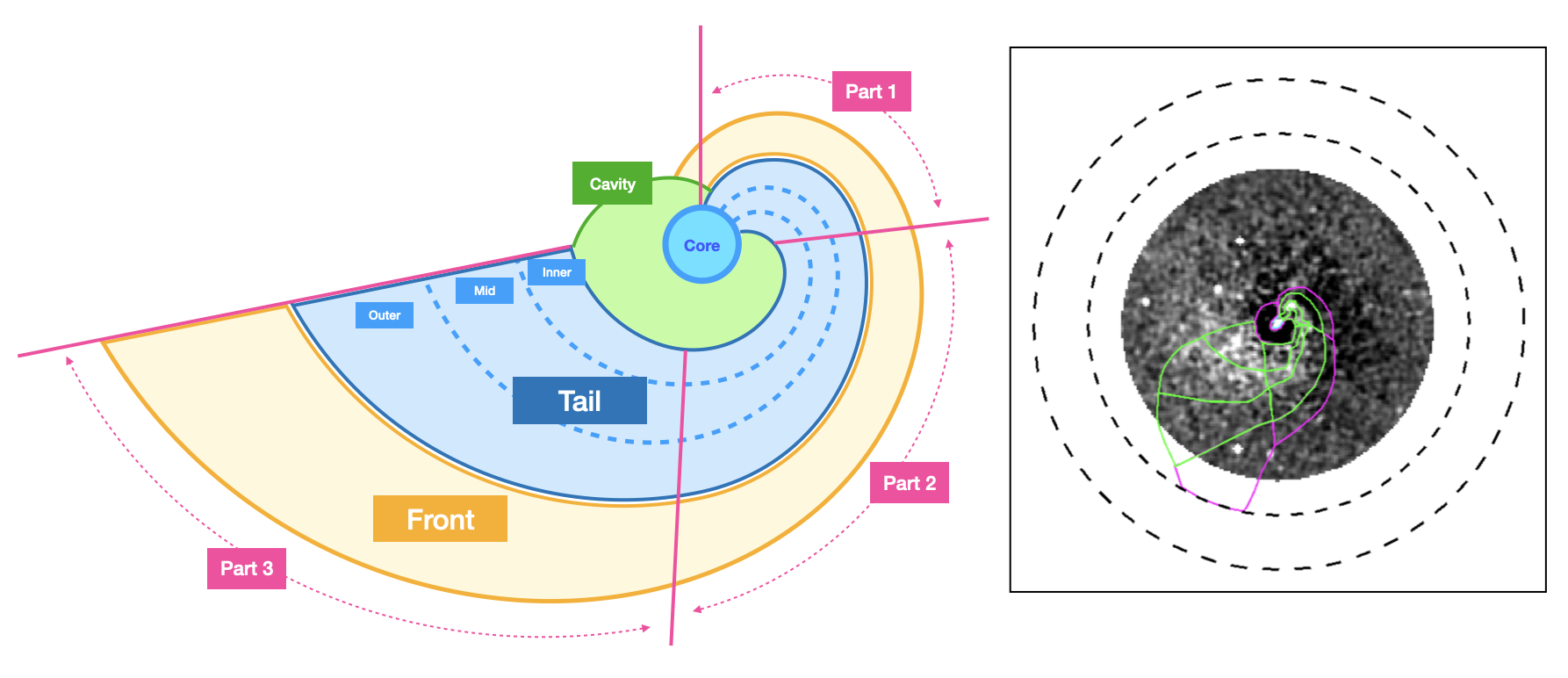}
   \caption{\textit{Left:} Schematic illustration of the regions defined for spatially resolved spectral analysis. The intracluster medium (ICM) is divided into four main sectors: Core (cyan), Cavity (green), Tail (blue-shaded spiral structure), and Front (yellow-shaded region). The Tail is further subdivided radially into In, Mid, and Out subregions, while the Front is divided angularly into Part 1, Part 2, and Part 3. \textit{Right:} Residual image of A2107 with the overlaid subregions.} 
   \label{fig:A2107-spiral-illustration}
\end{figure*}

\begin{figure}
   \centering
   \includegraphics[width=\hsize]{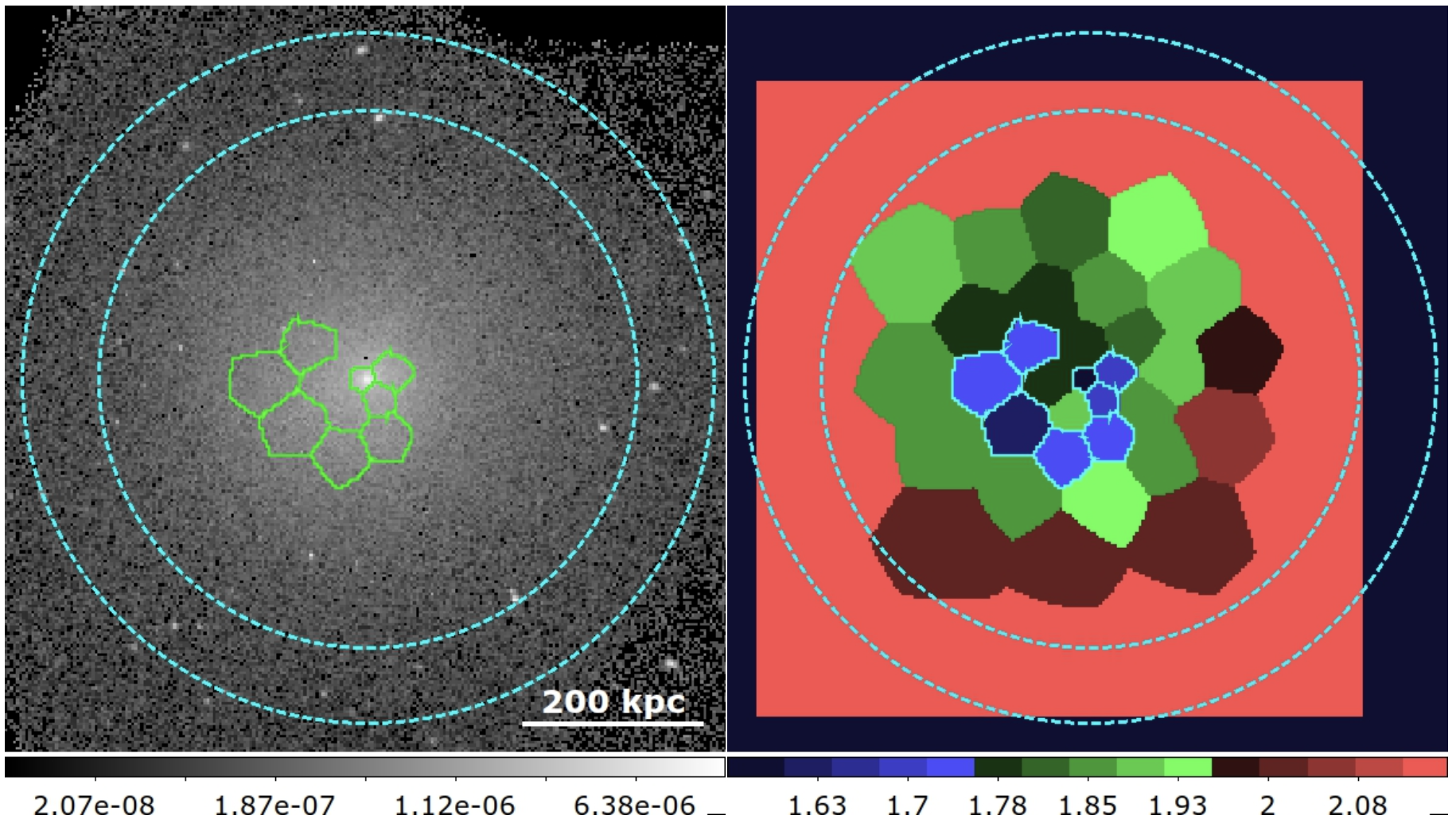}
   \caption{Merged X-ray image (\textit{left}) and pseudo-temperature map (\textit{right}) of A2107. Green WVT bins on the left panel indicate the relatively colder spiral feature.}
   \label{fig:A2107-xray-vorbin-regions}
\end{figure}

\begin{figure}
   \centering
   \includegraphics[width=\hsize]{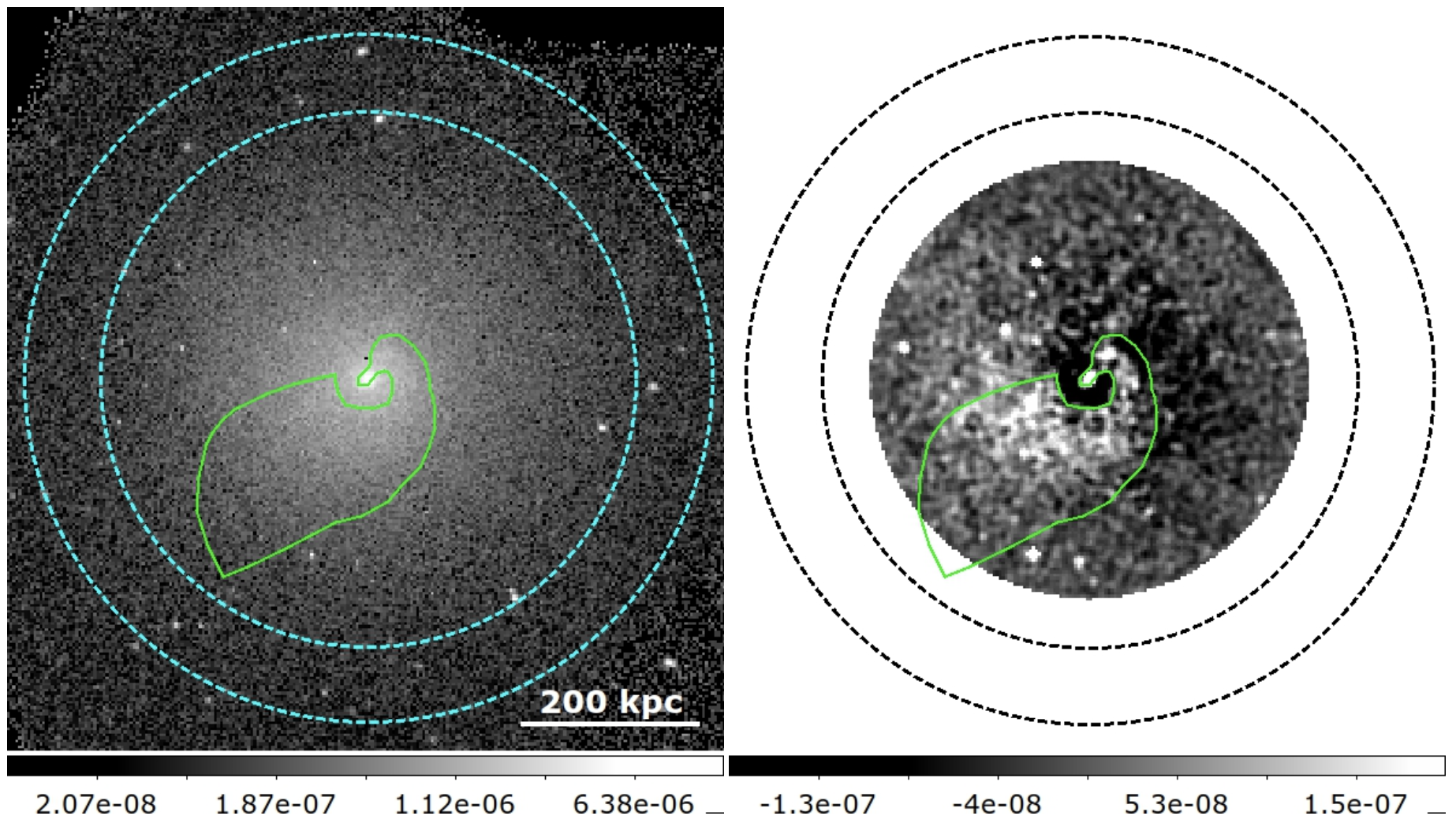}
   \caption{Merged X-ray image (\textit{left}) and residual image (\textit{right}) of A2107. The green region indicates the excess emission from the spiral feature.}
   \label{fig:A2107}
\end{figure}

\begin{figure*}
   \centering
   \includegraphics[width=\textwidth]{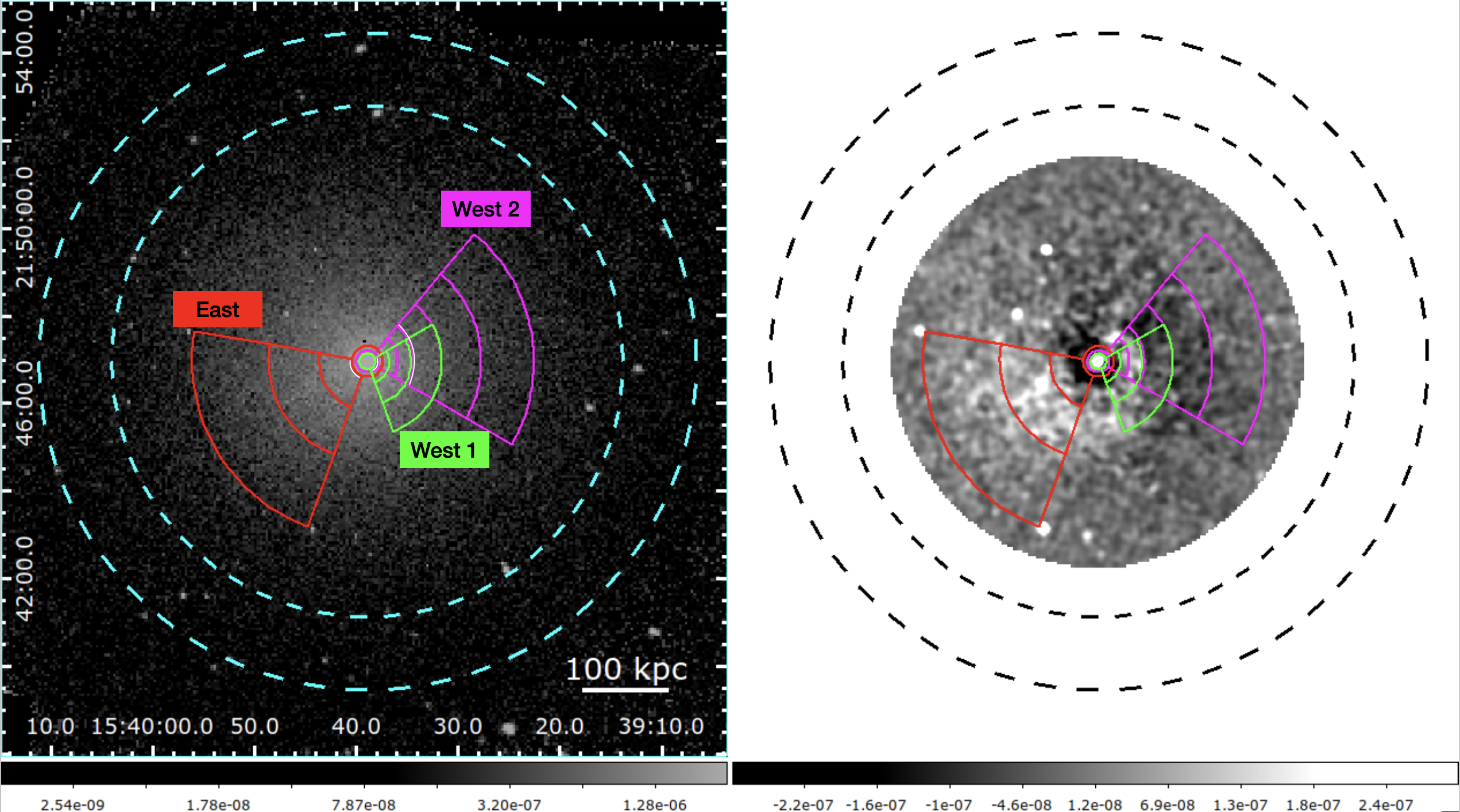}
   \caption{Merged X-ray image (\textit{left}) and residual image (\textit{right}) of A2107. West 1, East, and West 2 sectors are colored green, red, and magenta, respectively; cold front boundaries are marked with solid white arcs.}
   \label{fig:A2107-xray-cavity-panda-regions}
\end{figure*}

\begin{figure}
   \centering
   \includegraphics[width=\hsize, trim=10 20 10 30, clip]{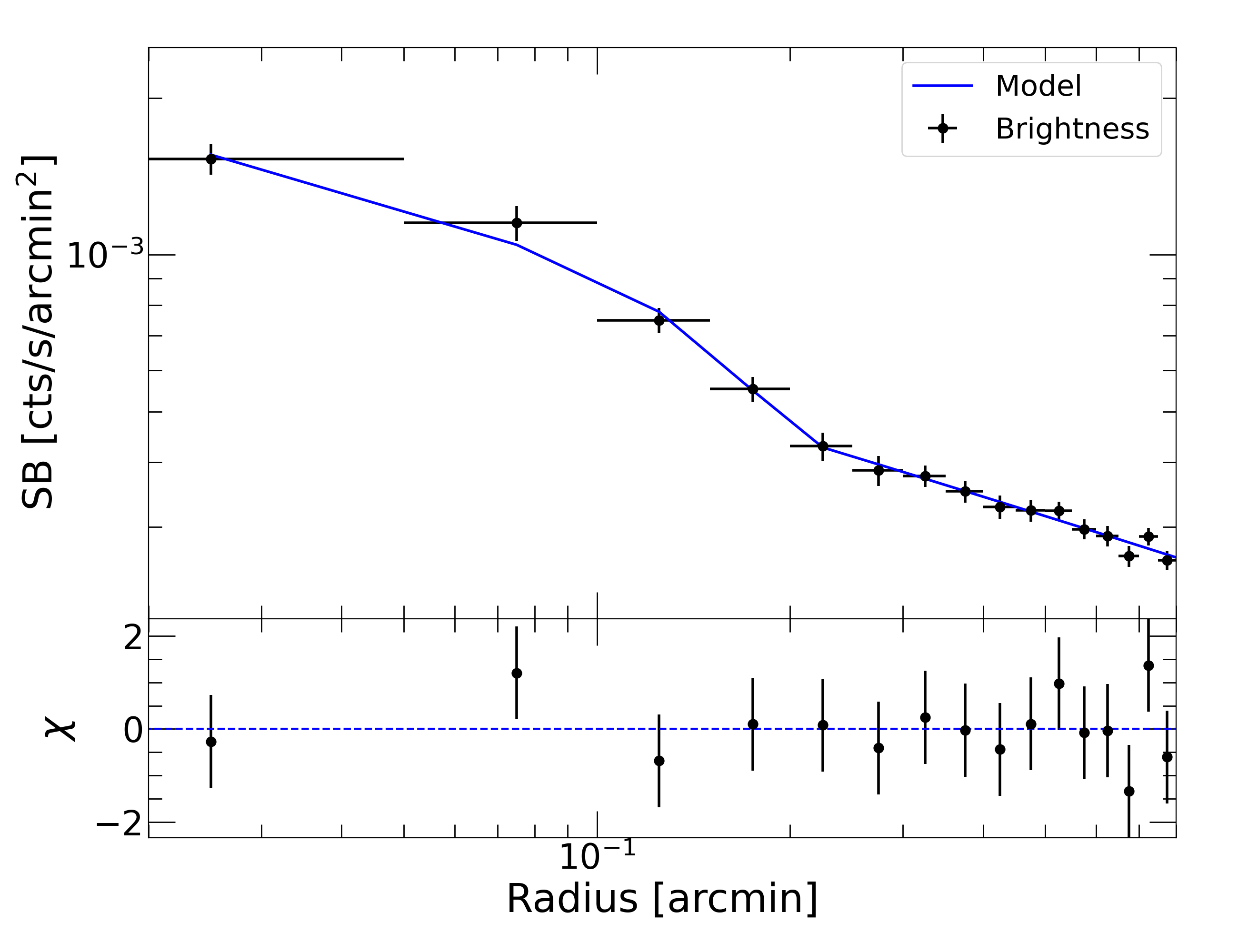}
   \includegraphics[width=\hsize, trim=10 20 10 30, clip]{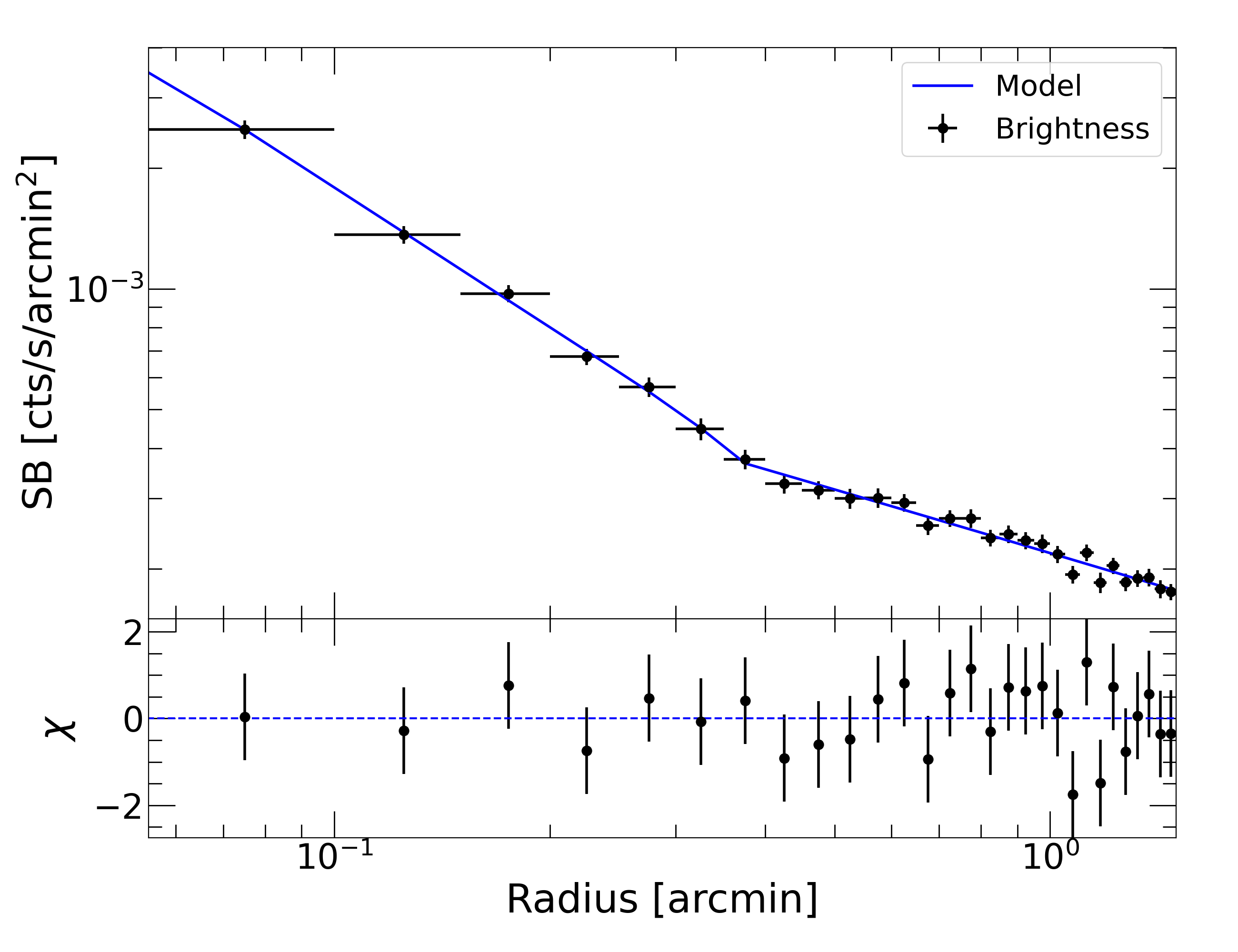}
   \includegraphics[width=\hsize, trim=0 20 10 30, clip]{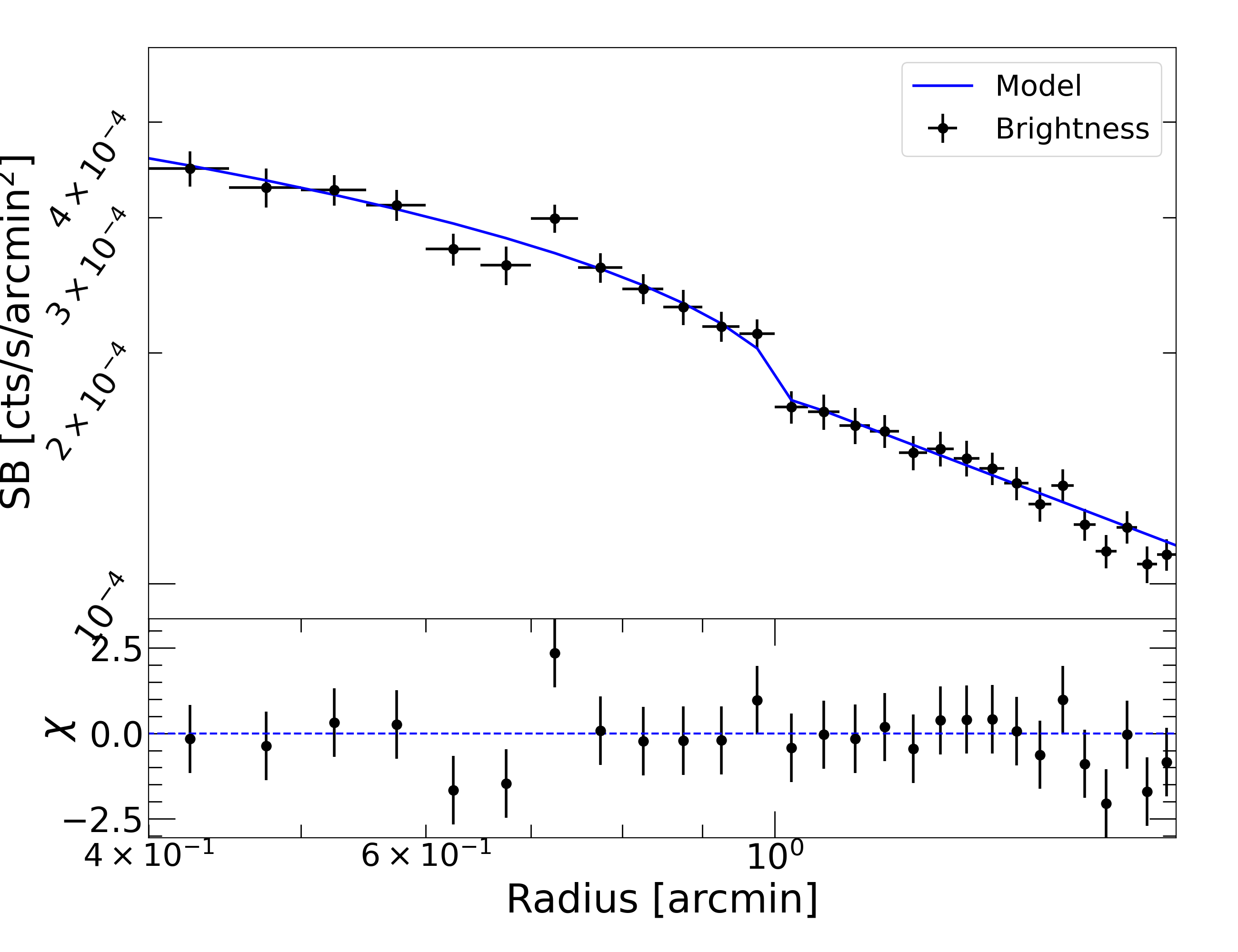}
   \caption{Surface brightness profiles of West 1 (upper), East (middle), and West 2 (lower) sectors, modeled with a broken-powerlaw. The break point of the model indicates the density discontinuity associated with the cold fronts.}
   \label{fig:A2107-panda-jump}
\end{figure}

\begin{figure}
   \centering
   \includegraphics[width=\hsize, trim=20 15 0 0, clip]{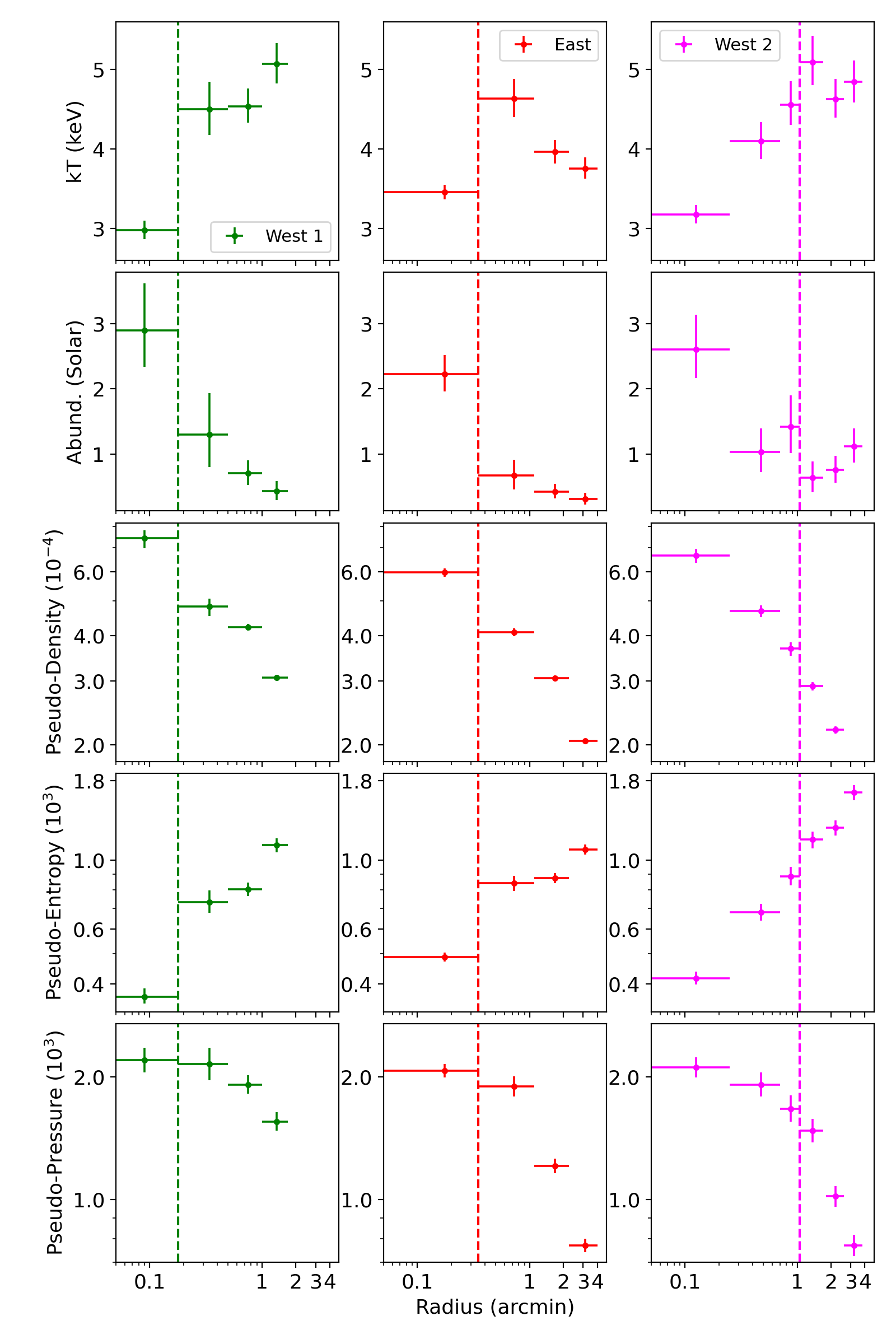}
   \caption{Radial profiles of the thermodynamic properties (temperature, metal abundance, pseudo-density, pseudo-pressure, and pseudo-entropy) extracted from West 1 (left), East (middle), and West 2 (right) sectors.}
   \label{fig:A2107-panda-profiles}
\end{figure}


\section{Results} \label{sec:results}

Here we present the analysis results for the Abell 2107 galaxy cluster, and the detection of a spiral-shaped substructure. We initially inspected the temperature structure of the ICM via a pseudo-temperature map (i.e. HR map). The pseudo-temperature map clearly revealed a spiral feature extending outward from the cluster center, colder than its surroundings (right panel of Figure \ref{fig:A2107-xray-cavity-vorbin}).

Additionally, we independently identified the same spiral structure in the surface brightness residual map, obtained by subtracting a double elliptical $\beta$-model from the X-ray image. In the residual map, the spiral appears as a region of excess emission, surrounded by a cavity (middle panel of Figure \ref{fig:A2107-xray-cavity-vorbin}).

\subsection{Defining the Boundaries of the Spiral}
To study the thermodynamic properties of the spiral, we adopted two different boundary-definition approaches and performed spectral analyses using both. In the first approach, we defined the spiral boundaries by selecting the colder Voronoi bins. To enable spectral comparison, we defined the bins immediately adjacent to these colder bins as the surrounding region of the spiral. The Voronoi bins used to define the spiral are shown in Figure \ref{fig:A2107-xray-vorbin-regions}. The fit results confirm that the spiral is colder than the surrounding plasma. It also exhibits higher abundance, density, and pressure, along with lower entropy. These features are consistent with the proposed scenario. However, when we focus on the tail of the spiral and exclude the central bin corresponding to the cool, dense core of the cluster, the temperature and abundance contrasts between the tail and its environment become much weaker, while the other properties remain statistically significant within 1$\sigma$ uncertainties. The fit results are displayed in Table \ref{tab:vorbin-spiral}.

\begin{table}
   \centering
   \renewcommand{\arraystretch}{1.2} 
   \begin{tabular}{lcc}
      \hline \hline
                           & kT (keV)        & $\chi^2 / dof$  \\ \hline
      Spiral               & $3.90 \pm 0.06$ & $1193.0 / 1142$ \\ \hline
      Spiral (center exc.) & $4.23 \pm 0.08$ & $1019.1 / 1035$ \\ \hline
      Surrounding          & $4.38 \pm 0.08$ & $1427.5 / 1485$ \\ \hline
   \end{tabular}
   \caption{Spectral temperature values of the spiral feature and the region surrounds it, defined using Voronoi binning. The spiral region was analyzed separately, both including and excluding the central bin. Shapes of the bins are shown in Figure \ref{fig:A2107-xray-vorbin-regions}}
   \label{tab:vorbin-spiral}
\end{table}

In the second approach, the spiral boundaries were defined based on the residual image, without being constrained by the Voronoi bin boundaries. This allowed for more flexible boundary definitions and enabled us to better trace the morphology of the spiral structure. Furthermore, by taking advantage of \textit{Chandra's} high spatial resolution, we were able to subdivide the spiral into smaller regions and investigate its internal structure in greater detail.

We examined the spiral pattern as two main components: the \textit{Core} and the \textit{Tail}. We further subdivided the tail along its length into \textit{Part 1}, \textit{Part 2} and \textit{Part 3}, and across its width into \textit{In}, \textit{Mid} and \textit{Out} regions. We also labeled the region surrounding the \textit{Core} as the \textit{Cavity} and the region surrounding the \textit{Tail} as the \textit{Front}. Note that the \textit{Front} region is also divided into three parts along its length, in the same way as the \textit{Tail}. This division, comprising a total of 14 regions, enabled a detailed analysis of the spiral's internal structure and its interaction with the surrounding plasma. The defined regions are presented in the right panel of Figure \ref{fig:A2107-spiral-illustration}, while a schematic illustration is shown in the left panel for clarity.

\begin{figure}
   \centering
   \includegraphics[width=\hsize]{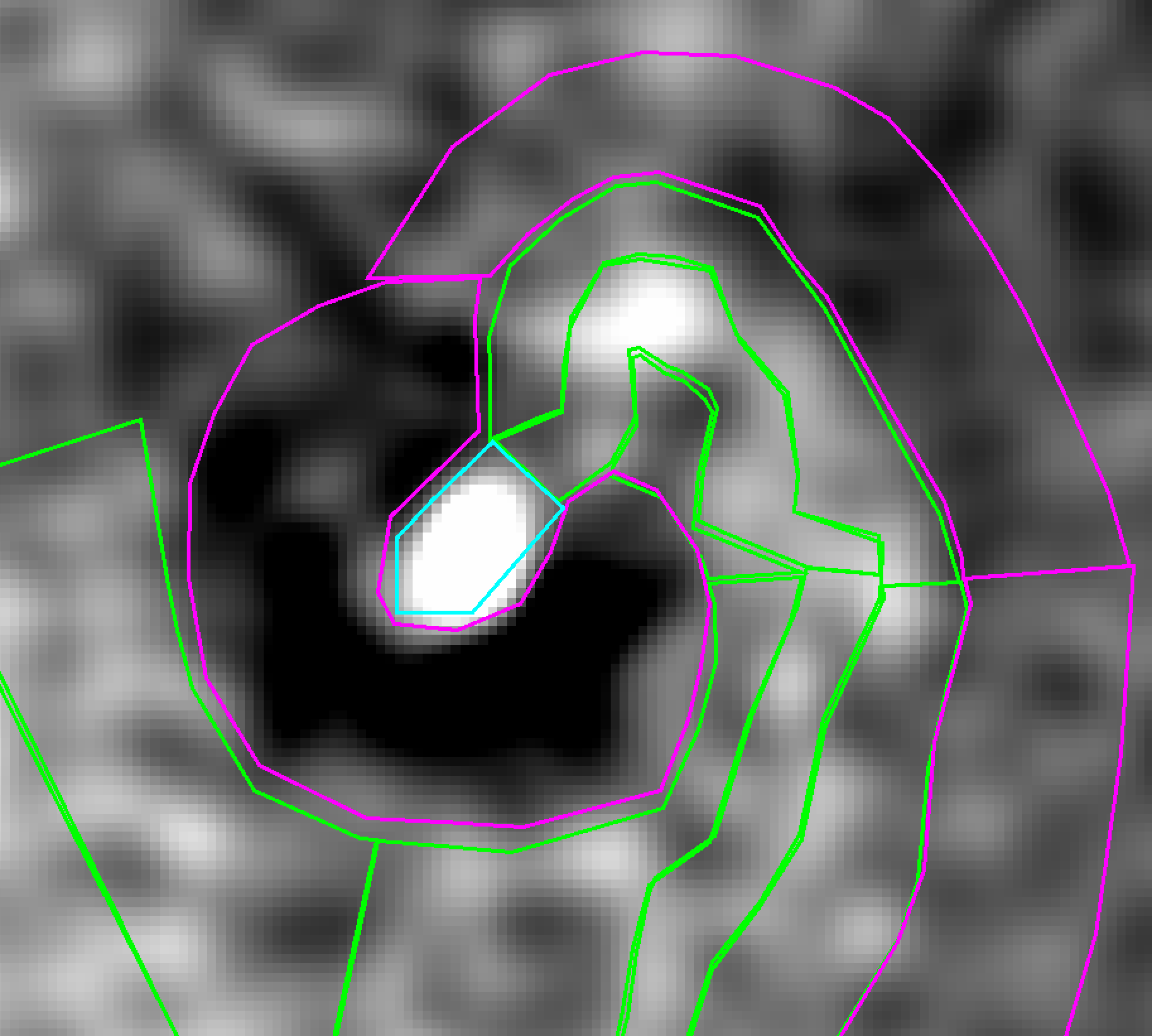}
   \caption{Central area of the residual image of A2107 with the overlaid subregions. (Zoomed-in view of the right panel of Figure \ref{fig:A2107-spiral-illustration}.)}
   \label{fig:A2107-spiral-zoom}
\end{figure}

In order to investigate the effects of turbulence and mixing both within the spiral and in the surrounding plasma, we aimed to trace the path of the coldest gas along the spiral when defining the region boundaries. Leveraging \textit{Chandra's} high spatial resolution and the long total exposure time, we carefully followed the coldest regions using an iterative approach in which region boundaries were manually adjusted to trace the coldest gas as closely as possible. As a result, we found that, particularly in the \textit{Mid} section of \textit{Part 1}, the coldest gas does not follow a straight path but instead traces a twisting path along the spiral. Zoomed image of this area is shown in Figure \ref{fig:A2107-spiral-zoom}.

\subsection{Thermodynamical Properties of the Spiral}
\begin{figure*}
   \centering
   \includegraphics[width=\textwidth, trim=60 75 70 80, clip]{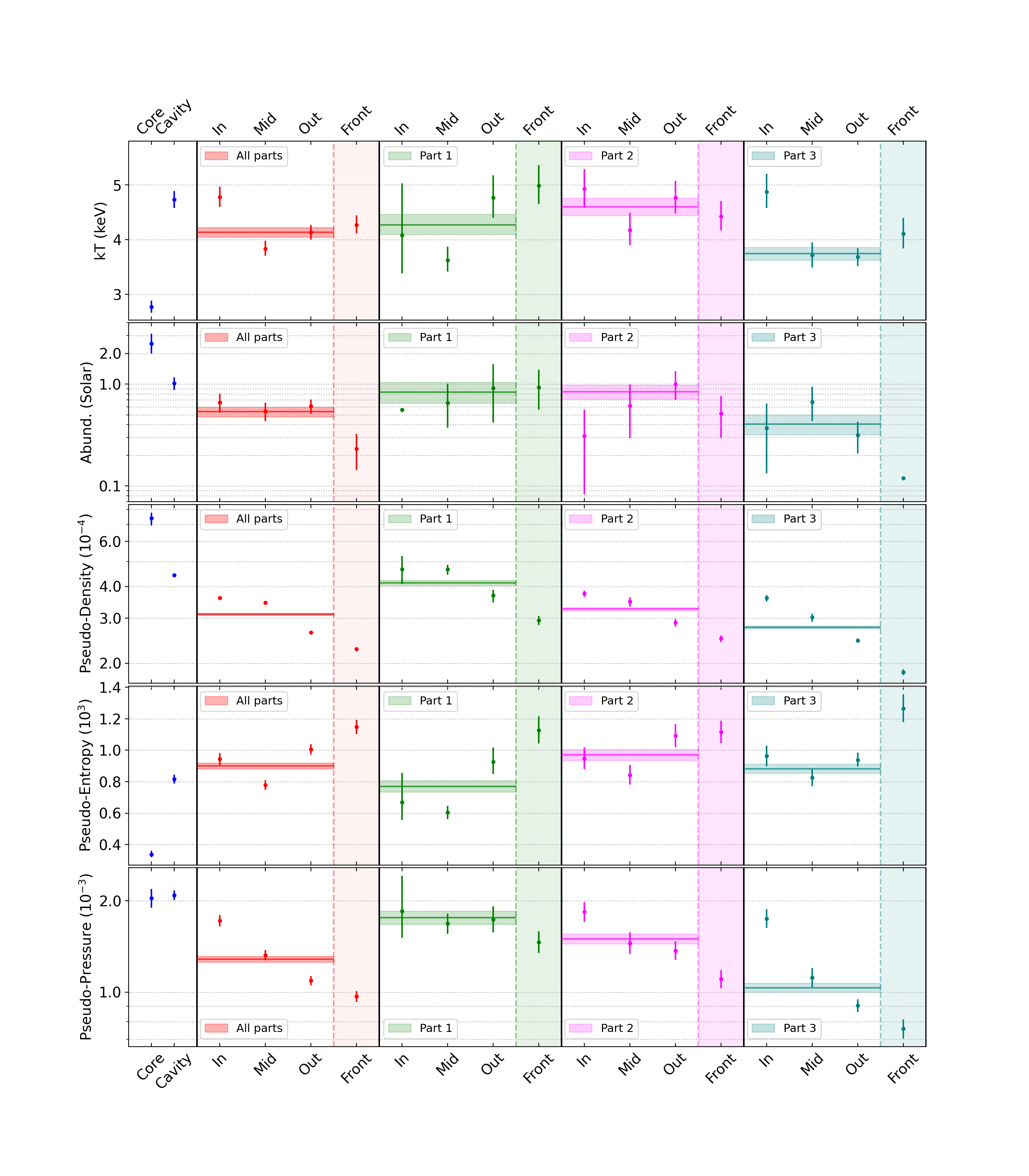}
   \caption{Spectrally derived parameters of the spiral feature and its surrounding regions, including all defined subregions.}
   \label{fig:A2107-spiral-profiles}
\end{figure*}

We present the thermodynamic properties of 14 regions that were described previously (Figure \ref{fig:A2107-spiral-profiles}). The spectral analysis shows that the \textit{Core} is characterized by low temperature, high metal abundance, low entropy, and relatively high pressure. In contrast, the \textit{Cavity} region is hotter, exhibits lower metal abundance and higher entropy, while its pressure remains approximately continuous with that of the \textit{Core}.

We first consider the \textit{Tail} as a whole. Its temperature is broadly consistent with that of the \textit{Front}. However, when divided into subregions, a more structured behavior emerges: the \textit{Mid} region is cooler, whereas the \textit{In} and \textit{Out} regions are relatively warmer. In terms of metallicity, both the overall \textit{Tail} and its subregions (\textit{In}, \textit{Mid} and \textit{Out}) are mutually consistent and exhibit higher values than the \textit{Front}. Density and pressure decrease from the \textit{In} regions outward and toward the \textit{Front}, with the exception that the \textit{In} and \textit{Mid} regions show comparable densiy values.

Among all quantities, entropy displays the most pronounced contrast between the \textit{Tail} and the \textit{Front}. Within the \textit{Tail}, the \textit{Mid} region has the lowest entropy, while the adjacent regions show slightly higher values, enhancing the overall contrast. This behavior persists throughout the \textit{Tail}. Additionally, we do not observe a monotonic trend in the thermodynamic properties across \textit{Part 1}, \textit{Part 2}, and \textit{Part 3}. 

In all segments, the \textit{Mid} subregions exhibit lower or comparable temperatures relative to the neighboring \textit{In} and \textit{Out} regions. A similar pattern is observed for entropy, with the \textit{Mid} subregions showing the lowest values in each segment, and maintaining a significant contrast with the \textit{Front}. In terms of abundance, the \textit{Tail} as a whole has a higher value than that of the \textit{Front} with $\sim2.8\sigma$ significance. However, subdividing the \textit{Tail} introduces large uncertainties, preventing firm conclusions about the inner structure at this level. The pressure profiles show similar behavior within the segments and for the \textit{Tail} as a whole, either remaining consistent or gradually decreasing from the \textit{In} regions toward the \textit{Front}.


\subsection{Cold Fronts}

We searched for density discontinuities at the edges of the spiral and modelled the surface brightness profiles using a broken-powerlaw model. The surface brightness profiles revealed three such discontinuities, which we confirmed as cold fronts by extracting thermodynamic profiles across these regions. The selected regions and the corresponding thermodynamic profiles are shown in Figure \ref{fig:A2107-xray-cavity-panda-regions} and Figure \ref{fig:A2107-panda-profiles}, respectively.

The first structure, located on the west side of the cluster center at a radius of $\sim9.3$ kpc (\textit{West 1}), shows sharp increases in temperature and entropy while the pressure remains continuous, indicating a cold front. This structure is positioned between the \textit{Core} and the \textit{Cavity} regions. The estimated density jump ratio for this front is $\mathcal{C}=1.71\pm0.18$.

The second cold front is located on the east side of the cluster center at a radius of $\sim17.9$ kpc (\textit{East}), shows the same thermodynamic behavior and is also situated between the \textit{Core} and the \textit{Cavity} regions. The estimated density jump ratio for this front is $\mathcal{C}=1.34\pm0.12$.

\begin{table*}
   \centering
   \renewcommand{\arraystretch}{1.2} 
   \begin{tabular}{cccccccc}
      \hline \hline
      Cold-Front & Sector Angles ($^\circ$)  & $\alpha_1$      & $\alpha_2$      & $r_f$ (kpc)           & $log(I_0)$       & $C$             & $\chi^2 / dof$ \\ \hline
      West 1 & $290 - 390$ & $0.58 \pm 0.08$ & $0.69 \pm 0.02$ & $9.2 \pm 0.5$ & $-2.64 \pm 0.09$ & $1.71 \pm 0.18$ & 7.44/11        \\ \hline
      East & $170 - 250$ & $1.14 \pm 0.06$ & $0.76 \pm 0.02$ & $17.9 \pm 2.0$ & $-3.11 \pm 0.14$ & $1.34 \pm 0.12$ & 16.7/24        \\ \hline
	  West 2 & $330 - 410$ & $0.39 \pm 0.08$ & $0.88 \pm 0.07$ & $52.7 \pm 1.0$ & $-3.79 \pm 0.03$ & $1.35 \pm 0.13$ & 13.0/14        \\ \hline \hline
   \end{tabular}
   \caption{Broken-powerlaw model fit results for the density discontinuities associated with the cold fronts, obtained using \texttt{pyproffit}. Here, $\alpha_1$ and $\alpha_2$ denotes the powerlaw indices before and after the jump radius $r_f$. $I_0$ is the central surface brightness value. $\mathcal{C}$ denotes the density jump ratio.}
   \label{tab:proffit}
\end{table*}

The third cold front is located on the west side at a radius of $\sim52.7$ kpc (\textit{West 2}), at the outer edge of the \textit{Part 1} section of the \textit{Tail}. In this case, the temperature transition is more gradual than in the inner fronts. The estimated density jump ratio is $\mathcal{C}=1.35\pm0.13$. A chip gap in one of the observations is located near this third cold front. To verify that the identified structure is not an artificial feature caused by this gap, the analysis was repeated excluding this observation. The results remain consistent ($\mathcal{C}=1.35\pm0.12$), confirming the robustness of the detection. The discontinuity fit results for cold fronts are presented in Figure \ref{fig:A2107-panda-jump} and Table \ref{tab:proffit}.

In terms of metallicity, all cold fronts exhibit higher values than the regions ahead of the fronts. The thermodynamic profiles are shown in Figure \ref{fig:A2107-panda-profiles}, where dashed vertical lines indicate the radial positions of the cold fronts.

\section{Discussion} \label{sec:discussion}

In this study, we investigated the thermodynamical properties of the sloshing spiral in A2107, focusing on its morphology and origin. Our analysis shows that the spiral arm exhibits lower temperature and entropy compared to the surrounding intracluster medium, while its metallicity is enhanced, indicating that it likely originates from sloshing cold gas uplifted from the cluster center. This is consistent with previous observations of other clusters, such as Perseus, Centaurus, and A2204 \citep{ghizzardi2014metal,fabian2011wide,sanders2006enrichment,sanders2005chandra,sanders2009giant}, where spiral structures also show enhanced metallicity and reduced temperature. Similar trends have been reported in recent studies of sloshing spirals \citep{zheng2025imaging,picquenot2025direct}, highlighting the robustness of these thermodynamic features.

\subsection{Internal Structure of the Spiral}

We subdivided the spiral into subregions to better understand its internal structure, local variations, and interaction with the surrounding medium, as described in the Section \ref{sec:results}. 

We find that the middle regions (\textit{Mid}) generally preserve their characteristics more clearly than the side regions (i.e., \textit{In} and \textit{Out}). This result shows that the side regions are more heavily affected by interaction and mixing with the ambient medium. In all three longitudinal segments (\textit{Parts 1, 2, and 3}), the coldest and lowest-entropy gas is consistently located in the \textit{Mid} subregions. Moreover, entropy shows the most pronounced contrast between the spiral arm and its surroundings, with comparatively smaller uncertainties. Therefore, among the examined thermodynamic quantities, entropy emerges as the most robust indicator for distinguishing the sloshing gas from the ambient environment. This consistent behavior along the longitudinal segments, are in line with previous findings that heat exchange and gas mixing are inefficient in the sloshing gas \citep{ghizzardi2014metal}, likely due to suppression by magnetic fields \citep{zuhone2011sloshing}. 

Another noteworthy result is that the physical properties of the \textit{Tail} do not vary monotonically along its length. This could be due to complex, spatially non-uniform plasma mixing, or to projection effects arising from the three-dimensional motion of the spiral, which may vary its inclination relative to the plane of the sky. As a result, gas located in front of or behind the moving arm may overlap along the line of sight, complicating the observed thermodynamic trends.

We detected three cold fronts at the boundaries of the spiral. They appear to have formed sequentially in opposite directions during the back-and-forth oscillation, and are arranged from the innermost to the outermost, roughly along the west, east, and again west directions. The innermost front exhibits the stongest density, temperature and entropy jumps. The second and third fronts have comparable density jumps; however, the second still shows a more pronounced contrast in temperature and entropy than the third. The weaker contrast may result from stronger ICM interaction in the outer spiral segments or partial disruption of the front structure, while projection effects may also contribute. 

Moreover, given the large extent of the spiral, additional cold fronts may exist but remain undetected. In particular, \textit{Front} region of the \textit{Part 3} exhibits a pronounced entropy jump (Figure \ref{fig:A2107-spiral-profiles}). This suggests the presence of an additional cold front, broadly consistent with the sequence and directions of the three detected cold fronts. However, we were unable to model a corresponding density jump. This may be due to the low photon counts in the region. Similar configurations have been reported in other clusters with sloshing spirals \citep{ghizzardi2014metal,kadam2024sloshing_1,zheng2025imaging}. In addition, numerical simulations by \cite{ascasibar2006origin} predict the formation of multiple cold fronts in opposite directions, and \cite{roediger2024toy} present a toy model describing such oscillatory motions of the cluster core.

In this context, several edge distortions along the spiral have been reported in previous studies and interpreted as signatures of Kelvin-Helmholtz instabilities (KHI). These features are observed at locations roughly perpendicular to the propagation direction of the cold fronts \citep{ghizzardi2014metal,zheng2025imaging}. Such structures are also predicted by numerical simulations \citep{zuhone2013cold,roediger2013kelvin}. A similar thinning is present in \textit{Part 2} of our system, at a location consistent with the geometric pattern reported in these studies, suggesting a possible association with KHI. Its spatial configuration is broadly consistent with a scenario in which magnetic fields only partially suppress KHI along cold front boundaries. This feature may therefore serve as a useful target for future high-resolution studies of ICM microphysics.


\subsection{Origin of the Spiral}

A2107 has presented a puzzling dynamical structure for a long time. \cite{oegerle1992structure} identified A2107 as a rotating cluster. This is also confirmed by subsequent studies \citep{kalinkov2005rotation,manolopoulou2016galaxy,liu2019testing}. Such large scale rotational motion is generally interpreted as evidence of past mergers or ongoing mass accretion from the surrounding environment.

In addition, high peculiar velocities have been reported for its cD galaxy \citep{girardi1997optical,oegerle2001dynamics}, providing further indication of merger activity. X-ray observations with \textit{Chandra} \citep{fujita2006chandra} revealed that the ICM surrounding the cD galaxy is compact and characterised by a short cooling time. The authors argued that maintaining such a configuration requires an additional heating source beyond AGN feedback, drawing a parallel with the Coma cluster. Finally they identified a pressure imbalance, which supports the notion that A2107 is not in a fully relaxed state.

On the other hand, the morphological evidence has shown no clear indication of a disturbed dynamical state. The galaxy velocity distribution in A2107 is strikingly Gaussian \citep{girardi1997optical}, and the cluster has been described as remarkably regular in its optical morphology \citep{kalinkov2005rotation}. Based on these analyses, \cite{kalinkov2005rotation} argued that the cluster's regular morphology does not support the presence of two overlapping structures (i.e., an ongoing merger). Additionally, X-ray analyses by \cite{lagana2010spiral} and \cite{song2018redshift} found no clear evidence of merger activity in the surface brightness of the cluster. \cite{song2018redshift} also noted the absence of X-ray cavities or radio bubbles, features typically expected in relaxed systems with AGN feedback.

Finally, \cite{cavagnolo2008entropy} found that the A2107's central entropy falls below the threshold expected to trigger cooling-induced star formation and $H_{\alpha}$ emission, yet no such emission is detected. This places A2107 in an anomalous category, further complicating its thermodynamic picture. Notably, A2029 exhibits a similar combination of a prominent spiral structure without any detectable perturber \citep{paterno2013deep} and low central entropy without clear cooling signatures \citep{cavagnolo2008entropy}, highlighting an interesting parallel between the two systems.

The detection of a sloshing spiral in A2107 strongly indicates an off-axis merger. Considering the findings of previous studies, such a merger could account for both the rotation and the peculiar velocity of the cD galaxy, thereby partially alleviating the existing tension. \cite{song2018redshift} detected at least five filamentary structures connected to known groups and clusters in the vicinity of A2107 and suggested that the observed rotation may originate from anisotropic galaxy infall along these filaments. However, in addition to the rotational motion, the presence of the spiral structure implies the existence of an infalling body with a non-zero impact parameter that transfers angular momentum to the cluster core. The absence of an optical or X-ray counterpart for such a perturber remains an open question.

While a merger-driven origin appears to provide the most natural explanation, alternative mechanisms have also been proposed for similar situations. For instance, \cite{zinger2018cold} presented a simplified one-dimensional model in which steady gas inflows toward the cluster core can induce dynamical disturbances in the ICM without requiring a merger event. If this mechanism plays a role in the dynamics of A2107, it could potentially be consistent with the detected filamentary structures \citep{song2018redshift}. Although based on simplified assumptions, this scenario provides a possible alternative to a merger-driven interpretation. Nevertheless, its relevance in realistic cluster environments is not yet well constrained, and a more definitive assessment will require improved simulations and higher-quality observational data. Future high-sensitivity X-ray observations may help clarify whether such large-scale inflows contribute to the observed ICM structure.

In this context, an off-axis interaction with a dark matter-only or dark matter-dominated subhalo represents a plausible scenario, consistent with both the required gravitational perturbation and the absence of any detectable optical or X-ray counterpart. However, the non-detection of a perturber alone does not uniquely establish this interpretation, and alternative scenarios involving an otherwise undetected perturber cannot be ruled out. In this framework, \cite{ascasibar2006origin} showed through simulations that the passage of a dark matter-only subhalo can perturb the cluster core and generate spiral-like cold fronts. Their model demonstrates that such an encounter induces a time-dependent gas velocity field in the ICM, producing ram pressure near the core and ultimately leading to the formation of cold fronts.

Additionally, \cite{powell2009relationship} found that nearly one-third of galaxy-sized dark matter substructures in their simulations lack identifiable X-ray counterparts, indicating that X-ray-dark subhalos can plausibly survive within clusters. Observationally, \cite{dupke2007different} detected a spiral arm with multiple cold fronts in A496, without any visible perturber, similar to A2107. They argued that their results are consistent with a merger involving a purely dark matter halo, whose gas may have been stripped in an earlier interaction.

From an observational perspective, the fact that this spiral structure was not detected in earlier studies of A2107 underscores the importance of deep exposures for revealing subtle ICM features. Similarly, in A2052, a spiral structure that was not detected in earlier observations was later revealed through longer exposures \citep{blanton2011very}. According to \cite{lagana2010spiral}, such spirals may be more common than previously assumed, and the identification of the spiral in A2107 suggests they could be even more frequent. Simulations have shown that ASTRO-H is capable of measuring the velocity signatures of sloshing motions \citep{zuhone2016simulating}. Upcoming high-sensitivity X-ray missions are expected to further advance this capability, making it possible to detect and study many more similar structures in other clusters in detail.

\section{Summary and Conclusions} \label{sec:conclusion}

In this work, we have presented a detailed thermodynamic investigation of the sloshing spiral structure discovered in the galaxy cluster Abell 2107, based on deep \textit{Chandra} X-ray observations. Through imaging and spectral analysis, we have characterized the morphology, internal structure, and cold front properties of this feature, and discussed its possible origin in the context of the cluster's dynamical history. Our main results are summarized as follows:

\begin{itemize}

   \item We report the discovery of a prominent sloshing spiral in the ICM of A2107, identified through pseudo-temperature map and $\beta$-model subtracted residual image, which had remained undetected in earlier, shallower observations.

   \item Spectral analysis shows that the spiral arm is significantly cooler and exhibits lower entropy and enhanced metallicity compared to the surrounding ICM, consistent with an origin in the cool, metal-enriched cluster core.

   \item Subregion analysis reveals that the central part of the spiral arm best preserves its intrinsic thermodynamic properties, while the outer regions show greater susceptibility to interaction and mixing with the ambient medium. Among the examined thermodynamic quantities, entropy provides the most robust contrast between the sloshing gas and its surroundings. Furthermore, the thermodynamic properties along the length of the spiral do not vary monotonically, likely reflecting spatially non-uniform mixing or projection effects arising from the three-dimensional motion of the spiral arm.

   \item We report three cold fronts distributed in nearly opposite directions along the same axis, consistent with numerical simulation predictions of off-axis merger-driven sloshing. A possible additional cold front in the outermost part of the spiral is suggested by a pronounced entropy jump, though low photon counts prevent definitive characterization.

   \item No optical or X-ray counterpart to a potential perturber is detected in A2107, a situation previously reported in other clusters such as A496 \citep{dupke2007different} and A2029 \citep{paterno2013deep}. An off-axis interaction with a dark matter-dominated or dark matter-only subhalo represents a plausible scenario for the observed sloshing features, broadly consistent with predictions from numerical simulations \citep{ascasibar2006origin,zuhone2010stirring}, though alternative scenarios involving an otherwise undetected perturber cannot be ruled out.
   
\end{itemize}

The detection of a sloshing spiral in A2107 provides new evidence for past dynamical activity in this apparently relaxed cluster, in line with the large-scale rotation and the peculiar velocity of the cD galaxy reported in previous studies. The identification of this structure further supports the notion that sloshing features may be more common than previously thought, and underscores the importance of deep X-ray exposures for revealing subtle ICM features. Upcoming high-sensitivity X-ray missions will be crucial for detecting and characterizing similar structures in a larger sample of clusters, and for directly measuring the gas velocity signatures associated with sloshing motions.

\section*{Acknowledgements}
We thank the anonymous referee for their constructive comments and suggestions, and we are also grateful to Turgay Çağlar, Onur Ünver, and Emine Gülmez for helpful discussions and valuable feedback on this work.

\section*{Data Availability}

The raw data used in this article are available to download at the \textit{Chandra X-ray Center} website\footnote{\url{https://cxc.harvard.edu/}}. The reduced data underlying this article will be shared on reasonable request to the corresponding author.



\bibliographystyle{mnras}
\bibliography{example} 








\bsp	
\label{lastpage}
\end{document}